# Universal thermodynamic framework for quasi-van der Waals epitaxy

Renhong Liang,[1] Mao Ye,[1] Renkui Zheng,[1] Jianhua Hao,[2] Haitao Huang,[2] Longlong Shu[3]*, Shanming Ke[1]*

[1]*School of Physics and Materials Science, Guangzhou University, Guangzhou 510006, China*

[2]*Department of Applied Physics, The Hong Kong Polytechnic University, Hong Kong, China*

[3]*School of Materials Science and Engineering, Nanchang University, Nanchang 330031, China*

**Abstract**
van der Waals (vdW) epitaxy is conventionally regarded as a rotation-free and strain-free growth mode driven by weak, isotropic interactions, yet many interfaces paradoxically exhibit strictly locked orientations that defy standard surface-energy models. We resolve this inconsistency by establishing a unified quantitative framework for 2D-3D systems, in which strong electrostatic and chemical interactions compete with entropic forces. We introduce a two-tier descriptor set—the predictive index ($I_{pre}$) and the thermodynamic locking criterion ($I_{lock}$)—to quantify the energetic sufficiency for locked epitaxy. Our theory accurately predicted the competitive interactions at the interface within the 2D-3D system, precisely characterized whether the epitaxial layer underwent free growth or was constrained in a locked growth mode, demonstrating robust consistency with diverse experimental observations. This framework unifies orientation selection in 3D-on-2D films and rotational locking in 2D-on-3D layers within a single-phase diagram. Our work provides a generalizable, predictive route to controlling epitaxial orientation across a broad spectrum of layered heterostructures.

**Teaser**
A unified framework resolves the competition between free and locked growth in 2D-3D heterostructures.

**Introduction**

Two-dimensional (2D) materials, featuring atomically flat and dangling-bond–free surfaces, have fundamentally transformed the landscape of heteroepitaxy (*1-4*). Their chemical inertness enables a growth mechanism classically termed van der Waals (vdW) epitaxy. In this idealized picture, the epitaxial layer mechanically decouples from the substrate. Weak, isotropic dispersive forces dominate interfacial interactions, allowing researchers to integrate diverse materials regardless of symmetry and lattice mismatch (*5,6*). While this paradigm holds for 2D-on-2D systems, integrating 2D layers with 3D bulk materials introduces a far more intricate landscape. The interface between a layered 2D material and a rigid 3D lattice acts not merely as a passive junction but as a critical boundary where symmetry breaking and distinct bonding natures collide (*7-9*). A comprehensive understanding of the 2D-3D interfaces—specifically regarding surface termination, polarity, and atomic-level reconstruction—is essential for effectively optimizing the electrical and optical properties of the resulting heterostructures (*10-12*).

However, we face significant challenges in achieving high-quality, orientation-controlled growth in such 2D-3D (called quasi–vdW) systems, particularly when depositing 3D nitrides or oxides on 2D substrates (*13,14*). Unlike the inert vdW gap, these interfaces often harbor dangling bonds or polar surfaces from the 3D side. These features induce complex reconstruction or chemical bonding, which fundamentally alters the epitaxial behavior. Consequently, experiments reveal that real 2D/3D and 3D/2D interfaces deviate sharply from the rotation-free and strain-free assumptions of classical vdW epitaxy. Instead of floating freely, epitaxial films often exhibit significant lattice strain and strong substrate locking (*15-17*). To describe these anomalies, researchers have proposed concepts such as hybrid vdW epitaxy (*18*), suggesting that a mixture of interactions governs the interface. These findings indicate a cognitive shift: the 2D-3D interface represents a complex arena where weak vdW force and non-vdW forces compete for dominance.

Two classes of long-standing paradoxes highlight how non-vdW forces—specifically electrostatic and chemical contributions—exert decisive control over epitaxial alignment. First, we observe strictly locked in-plane rotation in 2D/3D systems. Layered materials such as $MoS_2$ or graphene grown on polar 3D substrates (e.g., sapphire, GaN) frequently exhibit single-domain or strongly preferred in-plane orientations (*19-23*). This contradicts the rotational degeneracy that isotropic vdW forces should produce. Recent studies trace this locking to long-range electrostatic coupling and polarity-induced registry forces imposed by the substrate (*24-26*), which suppress rotational freedom and override the weak vdW potential (*27-29*). Second, anomalous orientation selection is observed in 3D/2D systems. In this inverse geometry, numerous reports show substrate-dependent selection of high-surface-energy orientations, such as (111)-oriented perovskites and (0001)-oriented wurtzite grown on mica and graphene (*15,16, 30-33*). These orientations defy the thermodynamic surface-energy–minimization criterion of vdW epitaxy. Such behaviors mentioned above further emphasize that the 2D-3D interface involves a competition among vdW force, long-range electrostatic forces, and chemical interactions. We propose that electrostatic pinning or strong, orientation-selective chemical bonding overrides the energetic penalty of forming unique preferential orientations or high-index facets.

Despite the growing recognition of these phenomena, the field lacks a unified understanding of 2D-3D quasi-vdW epitaxy. The coexistence of "free-floating" and "strongly locked" systems implies that interfacial coupling spans a broad spectrum. A fundamental gap remains that we lack a quantitative framework to predict how the competition between vdW force, electrostatic fields, and chemical affinity determines the growth modes and the interface features. Specifically, what physical principle dictates whether a given 2D-3D interface will behave as "free epitaxy" (dominated by vdW interaction) or "locked epitaxy" (dominated by electrostatic/chemical interaction)? To resolve this puzzle, we performed a systematic investigation combining controlled thin-film experiments with first-principles calculations. We identified orientation trends that classical theories cannot explain, which motivated us to develop a unified theoretical model that explicitly quantifies the multi-force complexity of the quasi-vdW interfaces.

**Results**
**Selective orientation epitaxial growth**

To examine how different 2D substrates influence the epitaxial orientation of 3D films, we first grew $SrTiO_3$ (STO) and $Fe_4N$ thin films on mica, $MoS_2$, and highly oriented pyrolytic graphite (HOPG), and systematically characterized their crystalline orientations by X-ray diffraction (XRD). As shown in Fig. 1A-C and Table 1, the STO films exhibit strikingly substrate-dependent out-of-plane orientations: (111) on mica, but (001) on both $MoS_2$ and HOPG. In sharp contrast, $Fe_4N$ displays the opposite trend—forming (111)-oriented films on $MoS_2$, yet adopting the (001) orientation on mica and HOPG (Fig. 1D–F).

Beyond out-of-plane orientation, the in-plane epitaxial relationships further reveal fundamentally different interfacial coupling behaviors. The pole figure of STO(111)/mica exhibits a clear six-fold symmetry (Fig. 1G), indicating that the STO lattice is azimuthally locked to the hexagonal surface symmetry of mica. In contrast, $Fe_4N$(001)/mica shows a 12-fold diffraction pattern (Fig. 1H), characteristic of multi-domain rotational variants. This behavior is reminiscent of well-known rotational degeneracy observed for 2D films on 3D substrates under weak vdW coupling (schematically illustrated in Fig. 1I). The coexistence of strong locking in STO/mica and rotational multiplicity in $Fe_4N$/mica suggests that the same 2D substrate can induce entirely different interfacial coupling regimes depending on the 3D material involved.

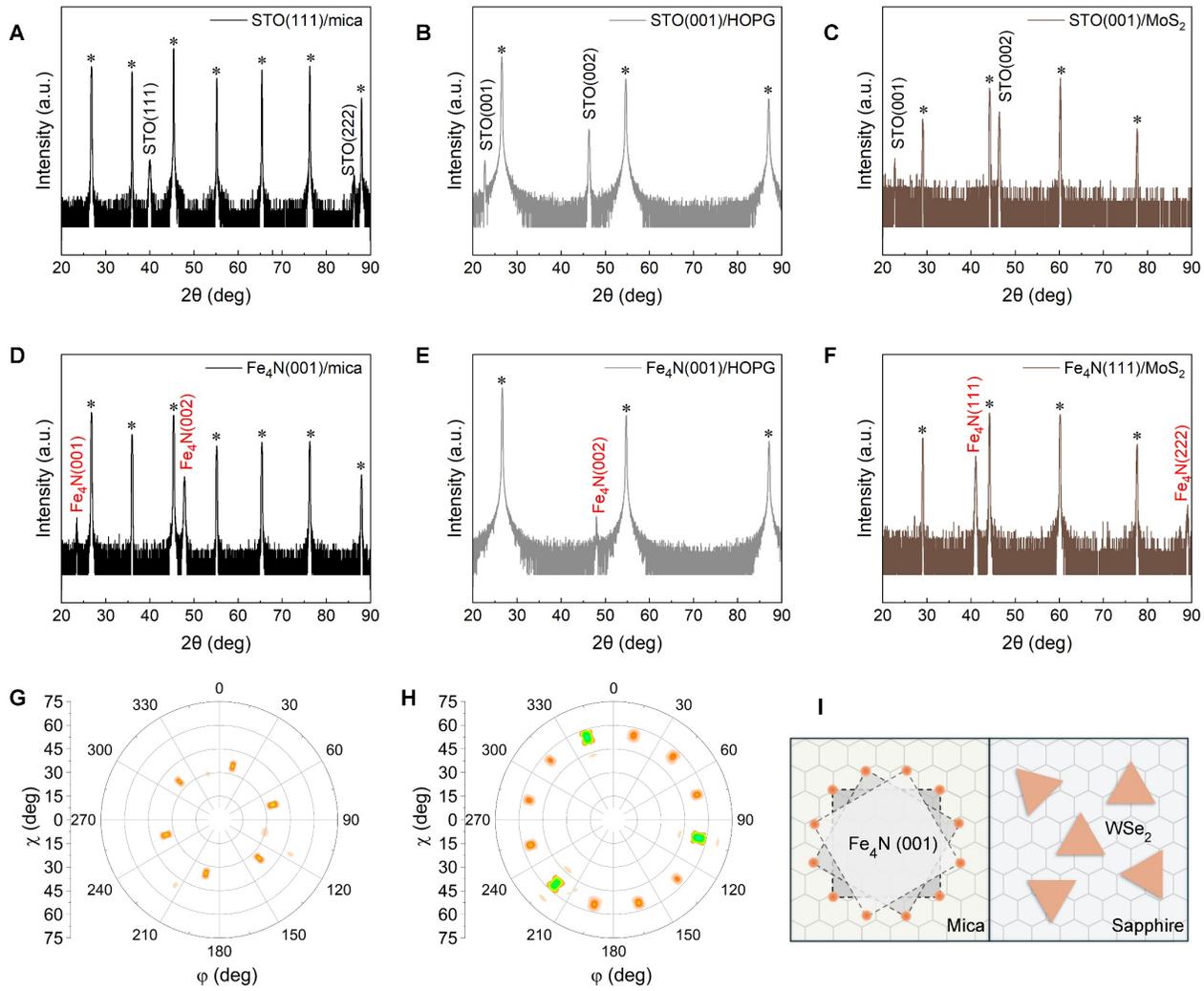

**Figure 1. Experimental evidence of orientation selectivity in 3D/2D systems.** (A-F) XRD θ-2θ scans of representative 3D/2D heterostructures showing substrate-dependent orientation selection: (A) STO(111)/mica, (B) STO(001)/MoS$_2$, (C) STO(001)/HOPG, (D) Fe$_4$N(001)/mica, (E) Fe$_4$N(001)/HOPG, and (F) Fe$_4$N(111)/MoS$_2$. XRD pole-figure measurements highlighting distinct in-plane epitaxial behaviors on mica: (G) STO(111)/mica exhibits 6-fold symmetry, indicating in-plane locking consistent with the mica surface. (H) Fe$_4$N(001)/mica exhibits 12 symmetry-related diffraction spots, characteristic of rotationally degenerate, weakly coupled vdW epitaxy. (I) Schematic illustration of the rotational degeneracy commonly observed in 2D films grown on 3D substrates, analogous to the rotational freedom manifested in Fe$_4$N(001)/mica.

A comparison of crystal structures provides further context (Fig. S1), both STO and Fe$_4$N adopt a cubic perovskite-type lattice in which the (001) surface is square and energetically the most stable, whereas the (111) facet forms a hexagonal layer but carries significantly higher surface energy. The 2D substrates—mica, MoS$_2$, and HOPG—possess hexagonal surface lattices (Fig. S2). Under the classical vdW assumption, where interfacial forces are dominated by weak, isotropic vdW interactions, 3D films should adopt their lowest-surface-energy orientation (001) and remain rotationally unconstrained (5). However, our experimental results deviate sharply from this prediction.

These observations expose a fundamental question at the heart of quasi-vdW epitaxy that what determines whether a 3D-2D interface behaves as a free epitaxy system dominated by surface energy, or a locked epitaxy system governed by strong interfacial electrostatic and chemical interactions? This puzzle motivates the need for a unified physical model capable of predicting when and why each regime emerges.

Table 1. Out-of-plane orientations of STO and Fe$_4$N epitaxial films on the 2D substrates

| Materials | Mica | HOPG | MoS$_2$ |
|---|---|---|---|
| SrTiO$_3$ | (111) | (001) | (001) |
| Fe$_4$N | (001) | (001) | (111) |

**Model**

To address the central question arising from the selective epitaxy in Fig. 1, we establish a quantitative framework grounded in the fundamental thermodynamic principle that epitaxial growth proceeds along the pathway that minimizes the total free energy. Within this view, the choice between free and locked epitaxy is determined by the competition between energetic costs and gains.

We define the thermodynamic driving force for epitaxial locking not as the absolute formation energy of a single interface, but as the relative free energy difference (ΔG) between the locked state and the thermodynamically preferred free state (reference state).

$$\Delta G = G_{locked} - G_{free} \approx \Delta\gamma_{sur} + \Delta E_{strain} + \Delta\gamma_{int} \quad (1)$$

here, each term represents a distinct energetic contribution and Δ denotes the energy difference between the locked state and the free state. $\Delta\gamma_{sur}$ is surface energy cost, represents the energy penalty arising from the formation of a thermodynamically unfavorable surface facet. For example, for a 3D-on-2D system, locking often requires exposing a high-energy facet to match the substrate symmetry, yielding $\Delta\gamma_{sur} > 0$. For a 2D-on-3D system, the film surface (basal plane) remains invariant under rotation. Thus, the surface energy contribution vanishes ($\Delta\gamma_{sur} \sim 0$), and the competition is governed solely by interfacial interactions. $\Delta E_{strain}$ represents the elastic strain energy penalty, which is significant for lattice-mismatched 3D films but negligible for vdW-bonded 2D layers. $\Delta\gamma_{int}$ is the difference in interfacial energy between the two states, arising from film-substrate interactions, this term is not a single entity but rather the sum of a weak, nonspecific vdW contribution and a strong, directional component characteristic of quasi–vdW epitaxy:

$$\Delta\gamma_{int} = \Delta\gamma_{vdW} + \Delta\gamma_{es} + \Delta\gamma_{chem} \quad (2)$$

here, $\Delta\gamma_{vdW}$ describes the isotropic vdW adhesion, whereas $\Delta\gamma_{es}$ and $\Delta\gamma_{chem}$ captures the electrostatic and chemical contributions that give rise to locking. Thus, selective epitaxy corresponds to the condition where the strong interfacial gain outweighs the total penalty.

Ideally, the growth mode could be determined by directly computing the total interfacial energy $\Delta\gamma_{int}$. However, calculating this value for every possible material combination and orientation requires computationally prohibitive supercell modeling. To overcome this bottleneck, we introduce a two-tier methodology that decouples rapid screening (Tier-1) from rigorous verification (Tier-2).

**Tier-1 (predictive): the interaction index $I_{pre}$**

The physical premise of our Tier-1 model is that the isotropic vdW component ($\Delta\gamma_{vdW}$) provides negligible directional guidance. Therefore, any "locked" force must originate exclusively from the anisotropic electrostatic ($\Delta\gamma_{es}$) and chemical ($\Delta\gamma_{chem}$) interactions. Accordingly, we formulated the interaction index $I_{pre}$ as a semi-empirical descriptor designed to rapidly estimate the propensity for quasi-vdW locking by assessing the polarity and chemical reactivity of the constituent surfaces without constructing the full interface. The index is defined as:

$$I_{pre} = \alpha P_{coupling} + \beta C_{affinity} \tag{3}$$

where $\alpha$ and $\beta$ are weighting factors, and $P_{coupling}$ serve as a proxy for the electrostatic contribution $\Delta\gamma_{es}$. According to the classical theory of intermolecular and surface forces, the electrostatic interaction energy between two polar entities scales linearly with the product of their respective dipole moments or surface field strengths (*34*). Therefore, we define the coupling index as the product of the polarities of the film and the substrate:

$$P_{coupling} = \Delta V_{film} \times \Delta V_{substrate} \tag{4}$$

here, the $\Delta V$ represents the normalized surface potential step, calculating by dividing the DFT-derived planar-averaged electrostatic potential *V(z)* by a reference energy of 1.00 eV. This normalization ensures dimensional consistency while preserving the magnitude of the electrostatic potential. The second term $C_{affinity}$ acts as a proxy for the chemical contribution ($\Delta\gamma_{chem}$). Crucially, for compound materials, we define the $C_{affinity}$ as the composition-weighted average of the adsorption energies of the constituent elements:

$$C_{affinity} = \sum x_i |E_{ads}^i| \tag{5}$$

here, $x_i$ represents the stoichiometric fraction of element $i$ (cations or anions) on the termination facet of the film, and $E_{ads}^i$ is the adsorption energy of the corresponding individual adatom on the substrate (also normalized by 1.00 eV). This definition is grounded in the surface thermodynamics of complex compounds. Taking STO as an example, it is widely confirmed that its surfaces manifest as stoichiometric terminations (e.g., $SrO/TiO_2$ or $SrO_3/Ti$) rather than pure elemental layers. Consequently, the interfacial adhesion is a collective property of the constituent species. The weighted approach captures this collective affinity, eliminating the bias of single-element descriptors and ensuring the model applies universally to both elemental and compound epilayers.

**Tier-2 (rigorous): the locking criterion $I_{lock}$**

$I_{lock}$ is a fully physical, DFT-based criterion with a sharp threshold at 1. Unlike the semi-empirical Tier-1 descriptor, $I_{lock}$ quantitatively evaluates whether quasi-vdW interactions are strong enough to dominate the relevant geometric degree of freedom. Because quasi-vdW manifests in two distinct physical mechanisms, the rigorous criterion naturally splits into two "prongs":

Prong 1—in-plane and out-of-plane orientation locking (3D/2D systems). We define the orientation-locking ratio as (interfacial energy gain)/(energy cost):

$$I_{lock(3D/2D)} = \frac{|\Delta\gamma_{es}+\Delta\gamma_{chem}|}{\Delta\gamma_{sur}+\Delta E_{strain}+\Delta\gamma_{vdW}} \tag{6}$$

where $\Delta$ denotes the energy difference between the locked state and the free state. Locking occurs if and only if $I_{lock(3D/2D)} > 1$.

Prong 2—in-plane rotational locking (2D/3D systems). Directional locking occurs if the quasi-vdW interaction is stronger than the rotational energy ripple in the vdW adhesion, we define the rotational locking ratio as:

$$I_{lock(2D/3D)} = \frac{|\Delta\gamma_{es}+\Delta\gamma_{chem}|}{|\Delta\gamma(\theta)|} \tag{7}$$

here the denominator $\Delta\gamma(\theta)$ is the calculated binding energy difference ($\gamma_{max}(\theta) - \gamma_{min}(\theta)$) as the 2D film is rotated. Similarly, locking occurs if $I_{lock(2D/3D)} > 1$.

**Locked vs. free epitaxial interfaces**

To quantify the electrostatic component central to our quasi-vdW framework, we first determined the intrinsic polarities of the 2D substrates using KPFM measurements (Fig. S3 and Fig. S4). Nanosheets of mica, MoS$_2$, and HOPG were exfoliated onto a common grounded ITO substrate to ensure a unified reference. The measured surface potentials reveal a clear hierarchy: mica exhibits a strongly positive potential (~+50 mV), while HOPG (~−80 mV) and bulk MoS$_2$ (~−180 mV) are both negative. This confirms that mica possesses a unique and substantially stronger polar surface compared to the nearly non-polar MoS$_2$ and HOPG (*35*).

The DFT calculated planar-averaged electrostatic potential *V(z)* profiles provide a definitive clarification. As shown in Fig. 2A-B, 10-layer HOPG and MoS$_2$ slabs exhibit perfectly symmetric *V(z)* profiles with identical vacuum levels on both sides, yielding *ΔV* ≈ 0 eV (0.006 eV for MoS$_2$ and 0.001 eV for HOPG). This confirms their intrinsically non-polar nature. In sharp contrast, mica shows a profoundly asymmetric *V(z)* and a giant potential step of *ΔV* = 6.063 eV (Fig. 2C), directly reflecting its strong surface dipole. A parallel analysis on the films shows that STO(001) is non-polar (*ΔV* = 0.001 eV, Fig. 2D), whereas STO(111) is strongly polar due to charged (SrO$_3$)$^{4-}$/Ti$^{4+}$ stacking (*ΔV* = 2.297 eV, Fig. 2E). Together, these results establish a complete polarity map showing that locked epitaxy (e.g., STO(111)/mica) corresponds to a polar-on-polar interface, while free epitaxy (e.g., STO(001)/HOPG) corresponds to non-polar-on-non-polar coupling.

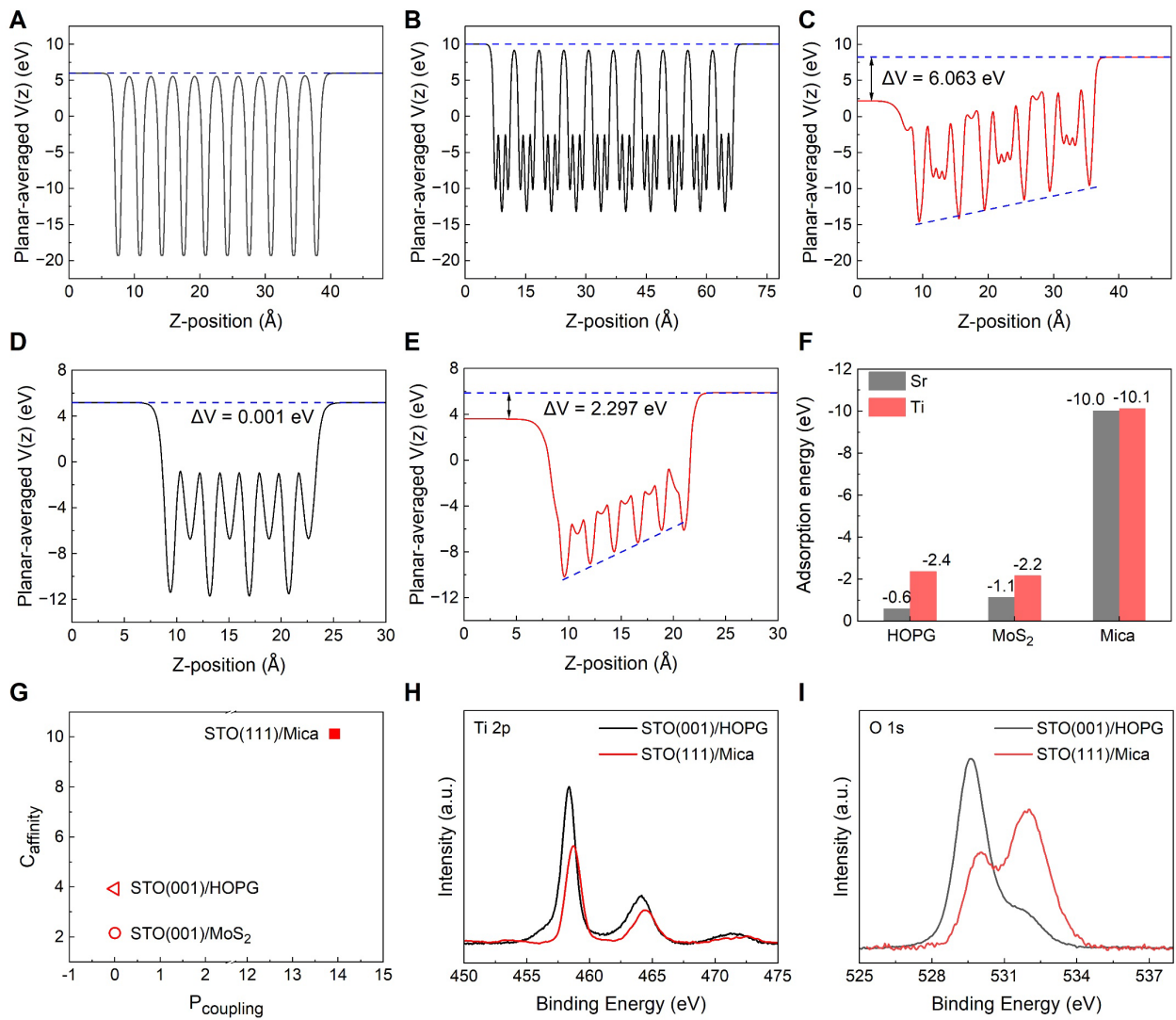

**Fig. 2. Quantification of intrinsic electrostatic polarity and chemical affinity determining the epitaxial driving forces.** (A-E) Calculated planar-averaged electrostatic potential profiles, *V(z)* for (A) HOPG, (B) MoS$_2$ (C) mica substrates, and (D) STO (001), (E) STO(111) films, highlighting the distinct polarity steps. (F) Calculated stable adsorption energies of Sr and Ti atoms on HOPG, MoS$_2$ and mica surfaces. (G) The calculated $C_{affinity}$ and $P_{coupling}$ for STO(111)/mica, STO(001)/HOPG, and STO (001)/MoS$_2$. (H) Ti 2p and (I) O 1s core-level XPS spectra comparing the STO(111)/mica and STO(001)/HOPG interfaces, showing binding energy shifts indicative of strong chemical/electrostatic coupling..

Turning to the chemical dimension, we calculated the adsorption energies of Sr and Ti adatoms to quantify the chemical affinity ($C_{affinity}$) (Fig. 2F). On the highly polar mica surface, both cations exhibit strong chemisorption at hollow sites ($E_{ads}$ > -10 eV). In contrast, adsorption on non-polar HOPG and MoS$_2$ is one to two orders of magnitude weaker, characteristic of physisorption. Integrating these electrostatic and chemical metrics into our model reveals that STO(111)/mica occupies a high-interaction regime in the $P_{coupling}$ vs. $C_{affinity}$ phase space, clearly separated from the low-interaction regimes of STO(001)/HOPG and STO(001)/MoS$_2$ (Fig. 2G). We experimentally verified the electronic consequences of this strong interaction using high-resolution X-ray photoelectron spectroscopy (XPS) (Fig. 2H-I and Fig. S5). Relative to the free STO(001)/HOPG reference, the core-level spectra (Sr 3$d$, Ti 2$p$, and O 1$s$) of locked STO(111)/mica consistently shift toward higher binding energies. This systematic shift signifies reduced valence electron screening arising from net electron depletion in the STO(111) film (*36*), confirming the significant interfacial charge transfer predicted by our polar-on-polar model.

To validate the universality of our framework beyond oxides, we extended our analysis to the cubic anti-perovskite nitride, Fe$_4$N (Fig. S6). Although Fe$_4$N consisting of charged planes similar to STO, its metallic nature induces strong electronic screening. Consequently, DFT calculations reveal that it retains a finite but significantly reduced polarity ($\Delta V$ = 0.135–0.251 eV) compared to the insulating STO. The chemical interaction landscape for Fe$_4$N is equally distinct, governed by the divergent behaviors of its cation (Fe) and anion (N) species. As shown in Fig. S6C, Fe adatoms exhibit strong chemisorption on polar mica surfaces (-5.20 eV) and, notably, essentially identical strong binding on MoS$_2$ (-5.20 eV) driven by robust Fe-S covalent bonding. However, the anion behavior breaks this symmetry, while nitrogen adatoms maintain high binding energies on HOPG and MoS$_2$ (about -5 eV), they exhibit significantly suppressed adsorption on the silicate-terminated mica surface (-1.50 eV), likely due to electrostatic repulsion between the anionic nitrogen and the oxygen-rich substrate. Incorporating these specific polarity and chemical affinity values into our phase diagram (Fig. S6D) reveals a distinct driving mechanism. Unlike STO(111)/mica, which is locked by electrostatics, the Fe$_4$N(111)/MoS$_2$ interface occupies the high-interaction locked regime primarily due to substantial chemical affinity ($C_{affinity}$) from both Fe and N sublattices, compensating for the reduced electrostatic coupling ($P_{coupling}$). Conversely, the reduced affinity of nitrogen on mica lowers the effective $C_{affinity}$ for Fe$_4$N(001)/mica, correctly positioning it in the free or transition regime consistent with the observed rotational disorder.

First-principles calculations of the planar-averaged charge density difference, $\Delta\rho(z)$, provide microscopic insight into these distinct growth regimes. For the STO(111)/mica polar-on-polar interface (Fig. 3A-B), substantial charge accumulation and depletion are observed at the junction regardless of termination (SrO$^{3-}$ or Ti$^{4+}$). This redistribution indicates the formation of a strong interfacial dipole, a hallmark of the locked regime driven by electrostatic continuity. Conversely, the non-polar-on-non-polar STO(001)/HOPG interface exhibits negligible charge redistribution (Fig. 3C-D), preserving the electronically inert nature of a pure vdW contact. Thus, the dramatic charge rearrangement in STO(111)/mica validates that the electrostatic coupling between the charged atomic planes of the film and the polar substrate is the fundamental driving force overcoming the vdW gap.

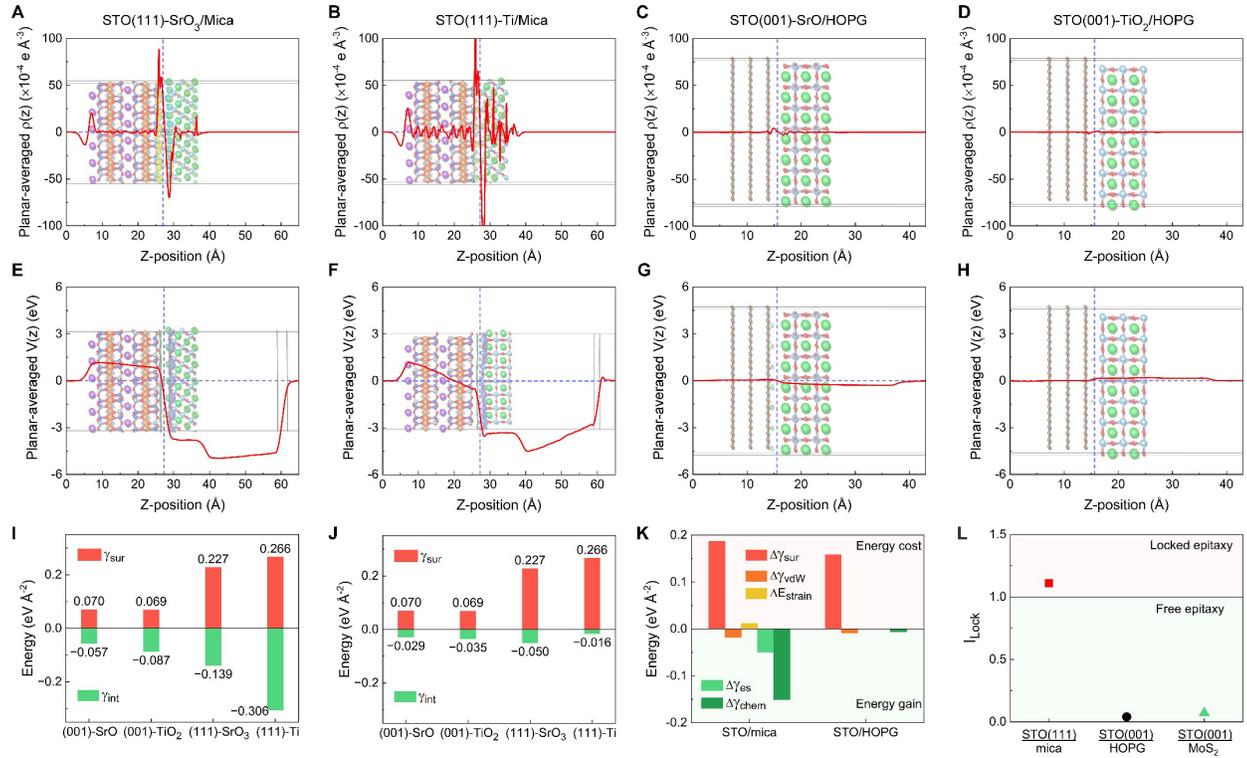

**Fig. 3. Thermodynamic verification of the locking criterion ($I_{lock}$) via interfacial charge transfer and energy competition analysis.** (A-D) Planar-averaged charge density difference $\Delta\rho(z)$ and (E-H) electrostatic potential profiles for the fully relaxed interfaces: (A, E) STO(111)-SrO$_3$/mica, (B, F) STO(111)-Ti/mica, (C, G) STO(001)-SrO/HOPG, (D, H) STO(001)-TiO$_2$/HOPG, respectively. Comparison of surface energy penalty $\gamma_{sur}$ and interfacial energy gain $\gamma_{int}$ for (I) STO/mica and (J) STO/HOPG heterostructures. (K) Comparison of energy cost ($\Delta\gamma_{sur} + \Delta\gamma_{vdW} + \Delta E_{strain}$) and energy gain ($\Delta\gamma_{es} + \Delta\gamma_{chem}$) for both systems. (L) The $I_{lock}$ descriptor for STO (111)/mica, STO (001)/HOPG, and STO (001)/MoS$_2$, when $I_{lock} > 1$ the locked epitaxy is occurred.

This difference in polarity is directly mirrored in the planar-averaged electrostatic potential profiles, *V(z)*. As shown in Fig. 3E-F, the STO(111)/mica interfaces exhibit a large potential step of ~4.9 eV across the interface, while the $V(z)$ profile for STO(001)/HOPG (Fig. 3G-H) is comparatively flat, showing almost no potential step, which confirms its non-polar, weakly interacting nature. The magnitude of this potential offset reflects the strength of interfacial electrostatic coupling and anticipates a substantial stabilizing contribution to the quasi-vdW interaction energy. To quantify this energetic contribution, we compared the surface energies $\gamma_{sur}$ of STO(001) and STO(111), as well as the interfacial energies $\gamma_{int}$ for STO/mica and STO/HOPG. Polar-on-polar STO(111)/mica yields a much larger $\gamma_{int}$ than either polar-on-nonpolar or nonpolar-on-nonpolar combinations, reflecting strong electrostatic and chemical stabilization. We further decomposed $\gamma_{int}$ and isolated the electrostatic component $\gamma_{es}$ using numerical integration of the charge-potential product, $\gamma_{es} = \frac{1}{2}\int \Delta\rho(z) \cdot V(z)dz$ (see Methods). For STO(111)/mica interface, $\gamma_{es}$ is found to be a large, negative value (-0.028 eV Å$^{-2}$ for SrO$_3$-termination and -0.05 eV Å$^{-2}$ for Ti-terminations), confirming strong electrostatic stabilization. However, $\gamma_{es}$ for STO(001)/HOPG is effectively zero (-3.1×10$^{-6}$ and -1.8×10$^{-7}$ eV Å$^{-2}$), verifying its vdW-only character.

Using Eqs. (1), (2), (3), and (6), we compared the total free-energy competition governing orientation selection. Fig. 3K shows that both STO/mica and STO/HOPG face a significant surface-energy penalty $\Delta\gamma_{sur}$ when forming the (111) facet, while $\Delta\gamma_{vdW}$ and $\Delta E_{strain}$ are comparatively small. Crucially, only the polar-on-polar STO(111)/mica interface provides a sufficiently large interfacial gain ($\Delta\gamma_{es}+\Delta\gamma_{chem}$) to overcome the high $\Delta\gamma_{film}$ penalty. In the nonpolar STO/HOPG system, the vdW-only $\Delta\gamma_{int}$ is far too weak to compensate for this cost. As a result, the $I_{lock}$ metric yields $I_{lock} > 1$ for STO(111)/mica, predicting locked epitaxy, whereas STO/HOPG and STO/MoS$_2$ (Fig. S7) exhibit $I_{lock} \ll 1$, corresponding to free epitaxy. Moreover, the similar analyses were performed on Fe$_4$N/2D systems (Fig. S8), showing $I_{lock} > 1$ for Fe$_4$N/MoS$_2$. These fully first-principles calculations provide direct validation of the energetic mechanism central to our quasi-vdW framework and are fully consistent with the experimentally observed orientation behaviors shown in Fig. 1.

For 2D-on-3D epitaxy, the in-plane rotational registry is governed by the competition between the interfacial locking potential and thermal fluctuations. We first examined the MoS$_2$/STO(001) interface as a reference for weak vdW coupling. As shown in Fig.4A and Fig. S9, the rotational energy landscape is extremely flat and featureless. Although it exhibits shallow minima at 15°/45°, the rotational barrier is merely ~ 1 meV. This thermodynamically negligible barrier allows for continuous rotational disorder, consistent with the experimentally observed multi-domain orientations (*37,38*) and the negligible planar-averaged charge density difference $\Delta\rho(z)$ (Fig. 4B). In sharp contrast, the MoS$_2$/Sapphire interface exhibits a strictly defined locking mechanism.

Unlike the ambiguous orientation on STO, our DFT calculations reveal a deep double-well potential with energy minima at 0°/60° (Fig. 4C and Fig. S10). The rotational barrier reaches ~ 69 meV, which is nearly two orders of magnitude higher than that of the STO system. This robust barrier originates from the strong electrostatic coupling between the S-sublattice and the cation-sublattice, as evidenced by the significant interfacial charge redistribution (Fig. 4D). Crucially, this deep double-well landscape explains the prevalent experimental observation of 0°/60° twin domains that the strong electrostatic force strictly confines the MoS$_2$ domains into these two discrete thermodynamic minima, preventing the random orientational spread seen on free surfaces (*39,40*). Thus, MoS$_2$/sapphire represents a canonical locked system ($I_{lock} > 1$, Fig. 4E-F), where interfacial electrostatics dictate the registry, whereas MoS$_2$/STO remains a free epitaxy governed by entropy.

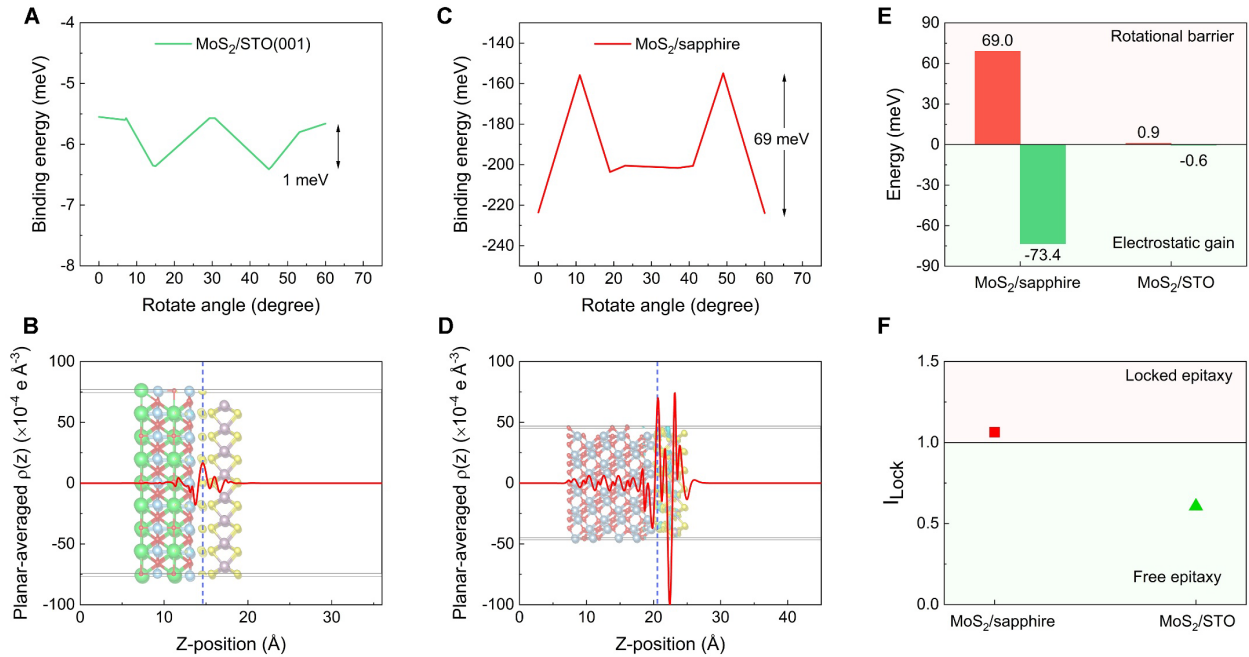

**Fig. 4. Mechanism of rotational locked vs. free rotation in 2D/3D systems.** (A, C) Calculated binding energy as a function of the relative orientation of (A) MoS$_2$/STO, and (C) MoS$_2$/sapphire. Note the magnitude difference in energy scales (1 meV vs. 69 meV). (B, D) Planar-averaged charge density difference $\Delta\rho(z)$ for (B) MoS$_2$/STO and (D) MoS$_2$/sapphire with a rotational angle of 15° and 0°, respectively. (E) Thermodynamic energy decomposition comparing the rotational barrier cost (defined as $\gamma_{max}(\theta) - \gamma_{min}(\theta)$) and the interfacial interaction gain. (F) The calculated rotational locking ratio for both systems. MoS$_2$/Sapphire falls into the "locked" regime driven by electrostatic forces, while MoS$_2$/STO remains in the "free" regime.

Having elucidated the locking mechanisms in cubic perovskite and chalcogenide systems, we sought to verify the universality of our model across a broader spectrum of material classes. We extended our DFT calculations to include wurtzite on mica (e.g., ZnO, GaN, Fig. S11), metals on dichalcogenides (e.g., Au/MoS$_2$, Fig. S12), graphene on metals (e.g., Gr/Cu, Fig. S13), and even 2D/2D systems (e.g., MoS$_2$/hBN, MoS$_2$/HOPG, Fig. S14). These systems encompass diverse bonding environments, ranging from ionic to metallic and to vdW interactions. To calibrate the weighting factors in our Tier-1 predictor, we mapped the growth modes of all investigated 2D-3D systems onto a $P_{coupling}$ vs. $C_{affinity}$ phase diagram (Fig. S15). A distinct linear boundary emerges, separating the experimentally observed locked systems from the free ones. By modeling this empirical phase boundary with the intercept form equation $P/P_0 + C/C_0 = 1$, where the critical intercepts are determined as $P_0 \approx 20$ and $C_0 \approx 5$, we derived the quantitative expression for the interaction index: $I_{pre} = P_{coupling} + 4C_{affinity}$. This derivation reveals a critical physical insight: the weighting factor of $\beta = 4$ for the chemical term implies that short-range chemical interactions are thermodynamically four times more potent in driving the locking transition than long-range electrostatic coupling per unit of dimensionless energy. This highlights the efficiency of local bonding in arresting rotational disorder.

Finally, we synthesized the data from all studied heterostructures into a master quasi-vdW epitaxy phase diagram (Fig. 5). This plot correlates the predictive potential ($I_{pre}$, Tier-1) with the rigorous thermodynamic locking criterion ($I_{lock}$, Tier-2), revealing a definitive bifurcation between the two growth regimes without outliers. Systems falling into the lower-left free epitaxy regime, such as STO/HOPG and MoS$_2$/STO, exhibit weak interfacial interactions ($I_{pre} < 20$) that fail to overcome thermal fluctuations or surface energy penalties ($I_{lock} < 1$), leading to rotation disorder or naturally favored low-energy facets. In contrast, the locked epitaxy regime comprises systems like STO/Mica, Fe$_4$N/MoS$_2$ and MoS$_2$/sapphire, where strong non-vdW forces—whether driven by electrostatics or chemical bonding—generate deep interfacial potential wells ($I_{lock} > 1$) that strictly enforce epitaxial alignment.

Crucially, the diagram establishes quantitative thresholds for engineered epitaxy that a minimum interaction of $I_{pre} \approx 20$ is required to activate the quasi-vdW mechanism, while the boundary at $I_{lock} = 1$ represents the thermodynamic limit where interfacial gains outweigh structural costs. This unified model resolves the long-standing contradictions in 2D-3D epitaxy, offering a predictive guideline for material synthesis by engineering interfaces to cross these boundaries, one can rationally design heterostructures with precise orientational control (*23, 38, 41-43*).

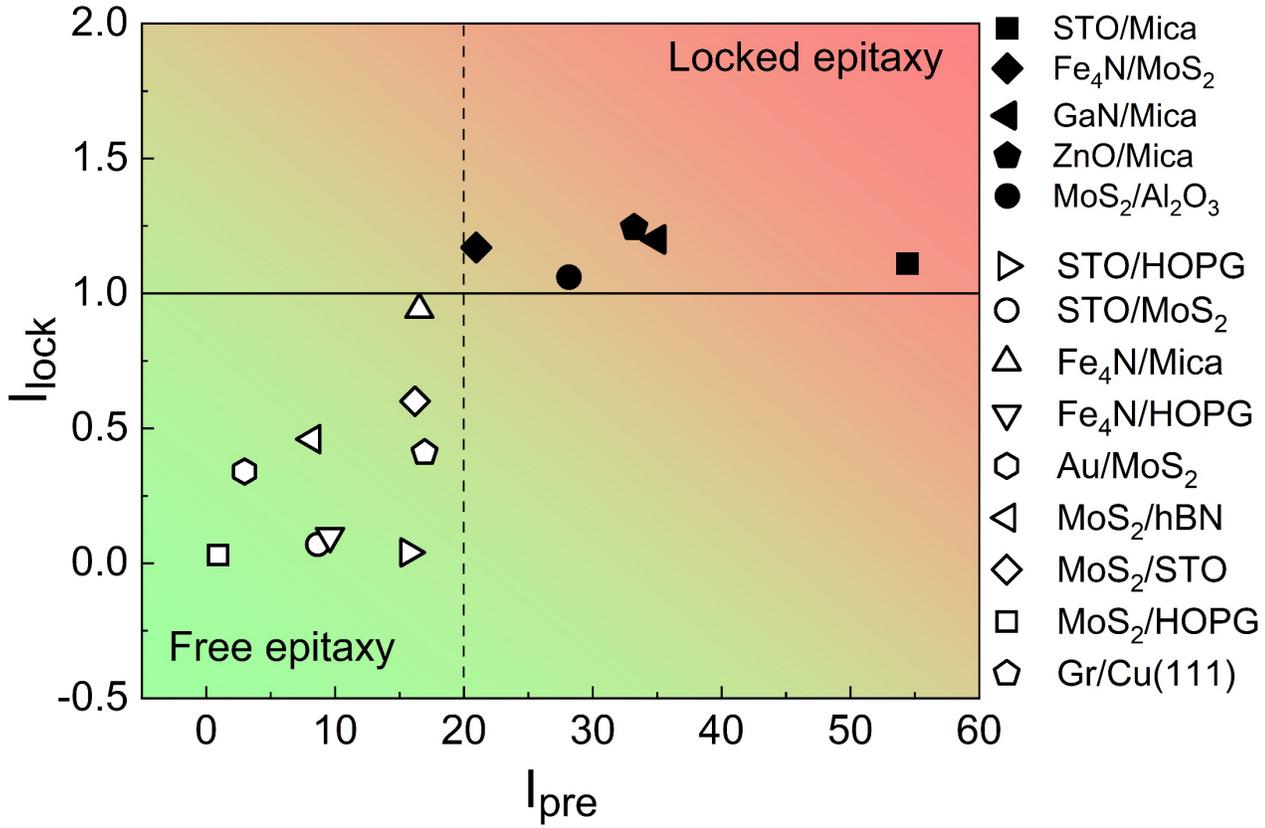

**Fig. 5. Universal phase diagram for quasi-vdW epitaxy.** The plot of the rigorous locking criterion ($I_{lock}$) versus the predictive index ($I_{pre}$) for various 2D/3D and 3D/2D heterostructures. The dashed vertical line at $I_{pre} \approx 20$ represents an empirical threshold for the interaction strength required to trigger locking. The solid horizontal line at $I_{Lock} = 1$ denotes the thermodynamic boundary where interfacial gains overcome the energetic costs of locking.

**Discussion**

The central achievement of this work is the establishment of a unified thermodynamic framework that rationalizes the diverse epitaxial selection rules observed in quasi-vdW heterostructures. By bridging the gap between 3D-on-2D and 2D-on-3D systems, we demonstrate that the transition between free and locked epitaxy is governed by a single, universal descriptor—$I_{lock}$. This criterion quantifies the competition between anisotropic interfacial coupling gains (driven by electrostatic and chemical potentials) and isotropic restoring forces (surface energy penalties or vdW ripples). Our framework reveals that these seemingly distinct growth modes are manifestations of the same physical principle: the system's drive to maximize interfacial overlap, provided the gain outweighs the energetic cost of symmetry adaptation.

A critical advance in our methodology is the use of DFT-calculated surface potentials ($\Delta V$) for the $P_{coupling}$ term, rather than macroscopic metrics like bulk ionicity. While bulk ionicity can rationalize general trends, it is fundamentally incapable of distinguishing between different crystallographic facets of the same material. The case of perovskite $SrTiO_3$ serves as a critical validation: its (001) surface is non-polar ($\Delta V \sim 0$), whereas its (111) surface is strongly polar. Our $\Delta V$ metric correctly captures this facet-dependent anisotropy, enabling the model to resolve the stark experimental contrast between free growth on STO(001) and locked growth on STO(111). Furthermore, our focus on the competition between (001) and (111) orientations for 3D-on-2D epitaxy is physically justified by symmetry considerations. The (001) facet offers the lowest intrinsic surface energy, driving the system toward a free state, while the (111) facet offers the only 6-fold symmetry match to hexagonal substrates, a prerequisite for locked epitaxy. Other low-index facets, such as (110) (2-fold symmetry), lack both the minimal energy of (001) and the symmetry compatibility of (111), rendering them thermodynamically uncompetitive.

Our framework's universality is effectively demonstrated by its ability to resolve the rotational alignment paradoxes in distinct interfaces. The $Fe_4N$(001)/mica system exemplifies a thermodynamically frustrated case. Despite a high interaction parameter ($P_{coupling}$ and $C_{affinity}$), the prohibitive thermodynamic cost of forming the polarity-matched (111) facet forces the film into the lower-energy (001) state. The resulting symmetry mismatch creates a degenerate energy landscape ($I_{lock} < 1$), leading to the observed rotational disorder. This stands in sharp contrast to the $MoS_2$/sapphire system, which our model identifies as a quintessentially locked system. The strong electrostatic coupling between the S-sublattice and the cation-sublattice generates a robust rotational barrier ($\sim 69$ meV), yielding $I_{lock} > 1$. This metric correctly predicts that the system is confined to discrete energy minima at 0° and 60°, explaining the prevalent observation of high-quality twin-domain epitaxy rather than continuous dispersion.

We acknowledge that our two-Tier model is fundamentally thermodynamic. It defines the energetic boundaries and the ultimate ground state for epitaxial selection but does not explicitly capture kinetic pathways. Kinetic factors—such as adatom mobility, deposition rate, and precursor supply—can play a decisive role, particularly for systems situated near the phase boundary ($I_{lock} \sim 1$). For instance, a high flux rate could kinetically trap a system in a metastable disordered state even if a locked orientation is thermodynamically favored (*44,45*). Future work integrating these kinetic barriers into the energy landscape will be essential for developing a complete, time-dependent description of quasi-vdW growth.

To apply these findings to the design of future heterostructures, the $I_{pre}$-$I_{lock}$ phase diagram provides a predictive roadmap. For applications requiring precise rotational alignment (e.g., Moiré quantum devices), researchers should engineer interfaces that cross the $I_{lock} = 1$ threshold. This can be achieved by selecting substrate-film pairs with high polarity matching or by modifying the substrate surface chemistry (e.g., creating specific terminations or step edges) to enhance the local electrostatic potential. Conversely, for applications benefiting from strain-free, rotationally compliant layers, selecting non-polar, low-affinity interfaces will promote the ideal vdW growth mode.

## Materials and Methods
### Thin film epitaxial growth

STO films were directly deposited on the freshly cleaved layered mica, $MoS_2$ and HOPG single crystal substrates by pulse laser deposition (PLD) using a KrF excimer laser ($\lambda$ = 248 nm). The layered materials we used in this study is purchased from Six Carbon technology (commercial product, Shenzhen, China). During growth, the temperature of the substrate was kept at 700 °C at an oxygen pressure of $1 \times 10^{-7}$ Torr, and the laser energy is ~1.5 J cm$^{-2}$, and a deposition frequency of 10 Hz. $Fe_4N$ films were deposited by DC magnetron sputtering (Kurt J. Lesker). The base pressure was lower than $1 \times 10^{-7}$ Torr. The substrate temperature was kept at 450 °C, an Ar (99.999%) and $N_2$ (99.999%) gas mixture with a mixing ratio of Ar:$N_2$ = 5:1 was introduced into the chamber. The total pressure was 8.5 mTorr, and the sputtering power on the target was 38 W.

### Characterizations

The crystal structure and epitaxial relationship of the films were measured by High resolution X-ray diffraction (XRD, Rigaku smartlab) with Cu K$\alpha$ radiation ($\lambda$ = 1.5406 nm). The surface potential were determined using atomic force microscopy (AFM, Cypher ES, Asylum Research) with Kelvin Probe Force Microscopy (KPFM) mode with a conductive tip cantilever (ASYELEC.01-R2). X-ray Photoelectron Spectroscopy (XPS) spectra with Ar-ion etching of samples are measured with a magnesium K$\alpha$ source (Thermo SCIENTIFIC K-ALPHA) after calibrating the peak with C 1s at 284.8 eV.

### First-principle calculations

Density functional theory (DFT) calculations were performed using the Quickstep module of the CP2K package, employing the mixed Gaussian and plane-wave (GPW) method (*46*). Core electrons were treated using norm-conserving Goedecker-Teter-Hutter (GTH) pseudopotentials (*47*). For structural relaxations, the Kohn-Sham orbitals were expanded using a DZVP-MOLOPT-GTH Gaussian basis set (*48*) with a plane-wave cutoff of 400 Ry. To ensure high accuracy in energy evaluations, a more complete TZV2P-MOLOPT-GTH basis set was employed, combined with an increased plane-wave cutoff of 600 Ry. The exchange-correlation interaction was described by the generalized gradient approximation (GGA) using the Perdew-Burke-Ernzerhof (PBE) functional (*49*). Long-range van der Waals dispersion corrections were applied using the DFT-D3(BJ) scheme (*50*) in all calculations. Geometry optimizations were converged until the maximum force and RMS force were below $4.5 \times 10^{-4}$ and $3.0 \times 10^{-4}$ Ha/Bohr, respectively, and the maximum atomic displacement was less than $3.0 \times 10^{-3}$ Å. A tight numerical precision (EPS_DEFAULT = $1.0 \times 10^{-12}$) was applied in all calculations. For slab models, the Poisson equation was solved using the MT scheme with periodic boundary conditions in the in-plane directions.

Prior to interface modeling, the bulk structures of STO, Fe$_4$N, mica, MoS$_2$, and HOPG were fully optimized. The calculated lattice constants for the cubic phases were $a_{STO}$ = 3.910 Å and $a_{Fe_4N}$ = 3.774 Å. For the substrate materials, the optimized parameters were determined as: $a$ = 5.347 Å and $b$ = 9.277 Å for monoclinic mica; $a = b$ = 3.16 Å for MoS$_2$; and $a$ = 2.46 Å for HOPG. Based on these results, slab models with specific surface terminations were constructed.

To construct the 3D/2D heterostructures, we employed large supercells to minimize the lattice mismatch between the film and the substrate to less than 5%. For the cubic-(001)/2D systems, the supercells consisted of a five-layer $7 \times 7$ STO(001) or Fe$_4$N(001) slab supported on: (i) a bilayer $5 \times 3$ mica supercell; (ii) a bilayer $5\sqrt{3} \times 9$ MoS$_2$ supercell; and (iii) a tri-layer $6\sqrt{3} \times 11$ HOPG supercell. For the hexagonal-(111)/2D systems, we constructed supercells containing a seven-layer $5\sqrt{2} \times 5\sqrt{2}$ STO(111) slab (supercell dimension $a \sim 33.1$ Å, $\gamma = 120°$) or Fe$_4$N(111) slab ($a \sim 32.0$ Å, $\gamma = 120°$) interfaced with the corresponding lattice-matched mica, MoS$_2$, and HOPG substrates. Due to the large system size, the orbital transformation (OT) method was applied for minimization, and Brillouin zone integration was performed at the Γ-point for these large supercells.

To investigate the energetics of rotational epitaxy under periodic boundary conditions, we constructed commensurate supercells using the Coincident Site Lattice (CSL) approach (*23*). This method identifies superlattices that accommodate both the film and substrate with minimal strain for a given relative rotation angle θ. Taking MoS$_2$ on the surface of sapphire as a representative example, the supercell lattice vectors for the MoS$_2$ film ($\vec{V}_{MoS_2}$) and sapphire substrate ($\vec{V}_{sapphire}$) were defined as linear combinations of their respective primitive vectors:

$$\vec{V}_{MoS_2} = m_1 \vec{a}_{MoS_2} + m_2 \vec{b}_{MoS_2} \tag{8}$$

$$\vec{V}_{sapphire} = n_1 \vec{a}_{sapphire} + n_2 \vec{b}_{sapphire} \tag{9}$$

where $m_{1,2}$ and $n_{1,2}$ are integers, and $a$ and $b$ represent primitive lattice vectors. To ensure the physical validity of the models, we selected integer sets that yielded the desired rotation angles while maintaining a lattice mismatch ($\Delta\delta$) below 5%, defined as:

$$\Delta = \frac{|\vec{V}_{MoS_2}| - |\vec{V}_{sapphire}|}{|\vec{V}_{sapphire}|} \times 100\% \tag{10}$$

To calculated the surface energy of films, such as STO(001) and STO(111), we initially calculated the cleavage energy of STO, $E_{cleavage}$, and distributed it to different terminations. For example, the unrelaxed internal cleavage energy of (001) surface was calculated as (*51*)

$$E_{cleavage}^{(001)} = \frac{1}{4}\left[E_{slab}^{(001)-SrO} + E_{slab}^{(001)-TiO_2} - x\Delta E_{bulk}\right] \tag{11}$$

where $E_{slab}^{(001)-SrO}$ and $E_{slab}^{(001)-TiO_2}$ are the energies of the unrelaxed SrO-terminated and TiO$_2$-terminated STO(001) surfaces, respectively, $\Delta E_{bulk}$ represents the formation energy of a single bulk STO unit cell, and $x$ is the number of total unit cells in both slabs. The factor of $\frac{1}{4}$ is included as four surfaces were created upon cleavage.

Based on $E_{cleavage}^{(001)}$ and $\Delta E_{bulk}$, the surfaces energy of the SrO-terminated STO-(001) surface can be calculated as

$$E_{surface}^{(001)-SrO} = E_{cleavage}^{(001)} + \frac{1}{2}\Delta E_{relaxation}^{(001)-SrO} \tag{12}$$

where $\Delta E_{relaxation}^{(001)-SrO}$ is the energy released upon the relaxation of the slab.

The binding energy was calculated according to $E_{bind} = E_{film/substrate} - [E_{substrate} + E_{film}]$, where $E_{film/substrate}$ is the total energy of the heterostructure, $E_{substrate}$ and $E_{film}$ are the total energy of the substrate slab and film slab, respectively. Basis-set superposition errors (BSSE) were estimated via the counterpoise correction to improve accuracy.

In our model, the total interface energy ($\gamma_{int}$) is decomposed into chemical ($\gamma_{chem}$), electrostatic ($\gamma_{es}$), and van der Waals ($\gamma_{vdW}$) contributions. Here, $\gamma_{vdW}$ specifically refers to the non-local dispersion forces (as calculated by the DFT-D3 method), while $\gamma_{es}$ encapsulates all other classical electrostatic interactions, including orientation and induction forces, which are captured by the standard DFT functional. To quantify the $\gamma_{es}$ arising from charge rearrangement at the heterostructure interface, we analyzed the planar-averaged charge density difference and its corresponding electrostatic potential. Using Multiwfn program (*52, 53*), the planar-averaged charge density difference, $\Delta\rho(z)$, was calculated and analyzed by subtracting the charge densities of the isolated components from the total charge density of the fully relaxed heterostructure:

$$\Delta\rho(z) = \rho_{total}(z) - \rho_{film}(z) - \rho_{substrate}(z) \tag{13}$$

The planar-averaged electrostatic potential, $V(z)$, was obtained by the similar method as that used for $\Delta\rho(z)$. The electrostatic energy per unit area ($\gamma_{es}$) stored in this interfacial charge layer is then given by the classical expression for the energy of a charge distribution in its own field (*54*)

$$\gamma_{es} = \frac{1}{2}\int \Delta\rho(z) \cdot V(z) dz \tag{14}$$

**References**

1. A. Koma, K. Sunouchi, T. Miyajima, Fabrication and characterization of heterostructures with subnanometer thickness. *Microelectron. Eng.* **2**, 129-136 (1984).
2. K. Ueno, T. Shimada, K. Saiki, A. Koma, Heteroepitaxial growth of layered transition metal dichalcogenides on sulfur- terminated GaAs{111} surfaces. *Appl. Phys. Lett.* **56**, 327-329 (1990).
3. A Koma, K Ueno, K Saiki, Heteroepitaxial growth by Van der Waals interaction in one-, two- and three-dimensional materials, *J. Cryst. Growth* **111**, 1029-1032 (1991).
4. A. Koma, van der Waals epitaxy—A new epitaxial growth method for a highly lattice-mismatched system. *Thin Solid Films* **216**, 72-76 (1992).
5. A. K. Geim, I. V. Grigorieva. Van der Waals heterostructures. *Nature* **499**, 419-425 (2013).
6. Y. Chu. Van der Waals oxide heteroepitaxy, *npj Quantum Mater.* **2**, 67 (2017).
7. A. Koma, K. Saiki, Y. Sato, heteroepitaxy of a two- dimensional material on a three-dimensional material. *Appl. Surf. Sci.* **42**, 451-456 (1989).
8. H. Sims, D. N. Leonard, A. Y. Birenbaum, Z. Ge, T. Berlijn, L. Li, V. R. Cooper, M. F. Chisholm, S. T. Pantelides, Intrinsic interfacial van der Waals monolayers and their effect on the high-temperature superconductor FeSe/SrTiO$_3$. *Phys. Rev. B* **100**, 144103 (2019).
9. S. Kim, S. Oh, S. J. Kwak, G. Noh, M. Choi, J. Lee, Y. Kim, M.-g. Kim, T. S. Kim, M.-k. Jo, W. B. Lee, J. Yoo, Y. J. Hong, S. Song, J. Y. Kwak, Y. Kim, H. Y. Jeong, K. Kang, Sequential multidimensional heteroepitaxy of chalcogen-sharing 3D ZnSe and 2D MoSe$_2$ with quasi van der Waals interface engineering. *Sci. Adv.* **11**, eads4573 (2025).
10. A. Mahata, E. Mosconi, Da. Meggiolaro, F. D. Angelis, Modulating band alignment in mixed dimensionality 3D/2D perovskites by surface termination ligand engineering. *Chem. Mater.* **32**, 105-113 (2020).
11. K. Reidy, G. Varnavides, J. D. Thomsen, A. Kumar, T. Pham, A. M. Blackburn, P. Anikeeva, P. Narang, J. M. LeBeau, F. M. Ross, Direct imaging and electronic structure modulation of moiré superlattices at the 2D/3D interface. *Nat Commun.* **12**, 1290 (2021).
12. Y. Meng, J. Feng, S. Han, Z. Xu, W. Mao, T. Zhang, J. S. Kim, I. Roh, Y. Zhao, D-H. Kim, Y. Yang, J-W. Lee, L. Yang, C-W, Qiu, S-H. Bae, Photonic van der Waals integration from 2D materials to 3D nanomembranes. *Nat. Rev. Mater.* **8**, 498-517 (2023).

13. K. S. Kim, J. Kwon, H. Ryu, C. Kim, H. Kim, E.-K. Lee, D. Lee, S. Seo, N. M. Han, J. M. Suh, J. Kim, M.-K. Song, S. Lee, M. Seol, J. Kim, The future of two-dimensional semiconductors beyond Moore's law. *Nat. Nanotechnol.* **19**, 895-906 (2024).
14. Y. Wang, S. Sarkar, h. Yan, M. chhowalla, Critical challenges in the development of electronics based on two- dimensional transition metal dichalcogenides. *Nat. Electron.* **7**, 638-645 (2024).
15. L. Lu, Y. Z. Dai, H. C. Du, M. Liu, J. Y. Wu, Y. Zhang, Z. S. Liang, S. Raza, D. W. Wang, C. L. Jia, Atomic scale understanding of the epitaxy of perovskite oxides on flexible mica substrate. *Adv. Mater. Interfaces* **7**, 1901265 (2020).
16. N. Wang, X. Pan, P. Wang, Y. Wang, H. He, Y-J. Zeng, L. Zhang, Y. Li, F. Wang, B. Lu, J. Huang, Z. Ye, Is all epitaxy on mica van der Waals epitaxy? *Mater. Today Nano* **20**, 100255 (2022).
17. K. Reidy, J. D. Thomsen, H. Y. Lee, V. Zarubin, Y. Yu, B. Wang, T. Pham, P. Periwal, F. M. Ross, Mechanisms of quasi van der Waals epitaxy of three-dimensional metallic nanoislands on suspended two-dimensional materials. *Nano Lett.* **22**, 5849-5858 (2022).
18. L. Hu, D. Liu, F. Zheng, X. Yang, Y. Yao, B. Shen, B. Huang, Hybrid van der Waals epitaxy. *Phys. Rev. Lett.* **133**, 046102 (2024).
19. L. Liu, T. Li, L. Ma, W. Li, S. Gao, W. Sun, R. Dong, X. Zou, D. Fan, L. Shao, C. Gu, N. Dai, Z. Yu, X. Chen, X. Tu, Y. Nie, P. Wang, J. Wang, Y. Shi, X. Wang, Uniform nucleation and epitaxy of bilayer molybdenum disulfide on sapphire. *Nature* **605**, 69-75 (2022).
20. X. Zou, Y. Zhao, D. Fan. S. Wu, Y. Wang, C. Zou, Y. Bian, L. Liu, L. Wu, Z. Han, W. Sun, Y. Nie, J. Gao, S. Zhu, Y. Shi, T. Li, F. Ding, X. Wang, Robust epitaxy of single-crystal transition-metal dichalcogenides on lanthanum-passivated sapphire. *Science* **390**, eaea0849 (2025).
21. J. Li, M. Chen, A. Samad, H. Dong, A. Ray, J. Zhang, X. Jiang, U. Schwingenschlogl, J. Domke, C. Chen, Y. Han, T. Fritz, R. S. Ruoff, B. Tian, X. Zhang, Wafer-scale single-crystal monolayer graphene grown on sapphire substrate. *Nat. Mater.* **21**, 740-747 (2022).
22. Z. Chen, C. Xie, w. wang, J. Zhao, B. Liu, J. Shan, X. wang, M. Hong, L. Lin, L. Huang, X. Lin, S. Yang, X. Gao, Y. Zhang, P. Gao, K. S. novoselov, J. Sun, Z. Liu, Direct growth ofwafer-scale highly oriented graphene on sapphire. *Sci. Adv.* **7**, eabk0115 (2021).
23. J. Dong, L. Zhang, X. Dai, F. Ding, The epitaxy of 2D materials growth. *Nat Commun.* **11**, 5862 (2020).
24. C. S. Chang, K. S. Kim, B.- i. Park, J. Choi, H. Kim, J. Jeong, M. Barone, n. Parker, S. Lee, X. Zhang, K. Lu, J. M. Suh, J. Kim, D. Lee, n. M. Han, M. Moon, Y. S. Lee, D.-H. Kim, D. G. Schlom, Y. J. Hong, J. Kim, Remote epitaxial interaction through graphene. *Sci. Adv.* **9**, eadj5379 (2023).

25. C. Liu, T. Liu, Z. Zhang, Z. Sun, G. Zhang, E. Wang, K. Liu, Understanding epitaxial growth of two-dimensional materials and their homostructures. *Nat. Nanotechnol.* **19**, 907-918 (2024).
26. R. Jia, Y. Xin, M. Potter, J. Jiang, Z. Wang, H. Ma, Z. Zhang, Z. Liang, L. Zhang, Z. Lu, R. Yang, S. Pendse, Y. Hu, K. Peng, Y. Meng, W. Bao, J. Liu, G-C. Wang, T-M. Lu, Y. Shi, H. Gao, J. Shi, Long-distance remote epitaxy. *Nature* **646**, 584-591 (2025).
27. X. Zhang, Z. Xu, L. Hui, J. Xin, F. Ding, How the orientation of graphene is determined during chemical vapor deposition growth. *J. Phys. Chem. Lett.* **3**, 2822-2827 (2012).
28. W. Kong, H. Li, K. Qiao, Y. Kim, K. Lee, Y. Nie, D. Lee, T. Osadchy, R. J. Molnar, D. K. Gaskill, R. L. Myers-Ward, K. M. Daniels, Y. Zhang, S. Sundram, Y. Yu, S.-H. Bae, S. Rajan, Y. Shao-Horn, K. Cho, A. Ougazzaden, J. C. Grossman, J. Kim, Polarity governs atomic interaction through two-dimensional materials. *Nat. Mater.* **17**, 999-1004 (2018).
29. P. Lacovig, M. Pozzo, D. Alfè, P. Vilmercati, A. Baraldi, S. Lizzit, Growth of Dome-shaped carbon nanoislands on Ir(111): the intermediate between carbidic clusters and quasi-free-standing graphene. *Phys. Rev. Lett.* **103**, 166101 (2009).
30. S. A. Lee, J. Hwang, E. S. Kim, S. W. Kim, W. S. Choi, Highly oriented $SrTiO_3$ thin film on graphene substrate. *ACS Appl. Mater. Interfaces* **9**, 3246-3250 (2017).
31. A. L. Bennett-Jackson, M. Falmbigl, K. Hantanasirisakul, Z. Gu, D. Imbrenda, A. V. Plokhikh, Al. Will-Cole, C. Hatter, L. Wu, B. Anasori, Y. Gogotsi, J. E. Spanier, van der Waals epitaxy of highly (111)-oriented $BaTiO_3$ on MXene. *Nanoscale* **11**, 622-630 (2019).
32. F. Ren, B. Liu, Z. Chen, Y. Yin, J. Sun, S. Zhang, B. Jiang, B. Liu, Z. Liu, J. Wang, M. Liang, G. Yuan, J. Yan, T. Wei, X. Yi, J. Wang, Y. Zhang, J. Li, P. Gao, Z. Liu, Z. Liu, Van der Waals epitaxy of nearly single-crystalline nitride films on amorphous graphene-glass wafer. *Sci. Adv.* **7**, eabf5011 (2021).
33. Y. Chang, W. Yang, W. Lo, Z. Lo, C. Ma, Y. Chu, Y. Chou, Direct growth of flexible GaN film via van der Waals epitaxy on mica. *Mater. Today Chem.* **26**, 101243 (2022).
34. J. N. Israelachvili, Intermolecular and Surface Forces, 3rd ed.; Academic Press: Burlington, MA, 2011.
35. H. K. Christenson, N. H. Thomson, The nature of the air-cleaved mica surface. *Surf. Sci. Rep.* **71**, 367-390 (2016).
36. Z. Zhong, P. Hansmann. Band alignment and charge transfer in complex oxide interfaces. *Phys. Rev. X* **7**, 011023 (2017).
37. M. K. Haastrup, M. Bianchi, L. Lammich, J. V. Lauritsen, The interface of in-situ grown single-layer epitaxial $MoS_2$ on $SrTiO_3$(001) and (111). *J. Phys.: Condens. Matter* **35**, 194001 (2023).

38. P. Chen, W. Xu, Y. Gao, J. H. Warner, M. R. Castell, Epitaxial growth of monolayer $MoS_2$ on $SrTiO_3$ single crystal substrates for applications in nanoelectronics. *ACS Appl. Nano Mater.* **1**, 6976-6988 (2018).
39. Y. Park, C. Ahn, J. Ahn, J. H. Kim, J. Jung, J. Oh, S. Ryu, S. Kim, S. C. Kim, T. Kim, H. Lim, Critical role of surface termination of sapphire substrates in crystallographic epitaxial growth of $MoS_2$ using inorganic molecular precursors. *ACS Nano* **17,** 1196-1205 (2023).
40. R. Ding, Z. Zhang, H. Wu, L. Deng, Y. Yuan, Y. Huang, M. Zhu, Z. Tian, Unidirectional epitaxy of wafer-scale $MoS_2$ on sapphire via growth kinetics control. *ACS Appl. Electron. Mater.* **7**, 8636-8645 (2025).
41. Chun-I Lu, C. Butler, J.-K. Huang, C. Hsing, H.-H. Yang, Y.-H. Chu, C.-H. Luo, Y.-C. Sun, S.-H. Hsu, K. O. Yang, C.-M. Wei, L.-J. Li, M.-T. Lin, Graphite edge controlled registration of monolayer $MoS_2$ crystal orientation. *Appl. Phys. Lett.* **106**, 181904 (2015)
42. L. Wang, X. Xu, L. Zhang, R. Qiao, M. Wu, Z. Wang, S. Zhang, J. Liang, Z. Zhang, Z. Zhang, W. Chen, X. Xie, J. Zong, Y. Shan, Y. Guo, M. Willinger, H. Wu, Q. Li, W. Wang, P. Gao, S. Wu, Y. Zhang, Y. Jiang, D. Yu, E. Wang, X. Bai, Z.-J. Wang, F. Ding, K. Liu, Epitaxial growth of a 100-square-centimetre single-crystal hexagonal boron nitride monolayer on copper. *Nature* **570**, 91-95 (2019).
43. L. Wang, J. Qi, W. Wei, M. Wu, Z. Zhang, X. Li, H. Sun, Q. Guo, M. Cao, Q. Wang, C. Zhao, Y. Sheng, Z. Liu, C. Liu, M. Wu, Z. Xu, W. Wang, H. Hong, P. Gao, M. Wu, Z.-J. Wang, X. Xu, E. Wang, F. Ding, X. Zheng, K. Liu, X. Bai, Bevel-edge epitaxy of ferroelectric rhombohedral boron nitride single crystal. *Nature* **629**, 74-79 (2024).
44. X. Zhang, T. H. Choudhury, M. Chubarov, Y. Xiang, B. Jariwala, F. Zhang, N. Alem, G.-C. Wang, J. A. Robinson, J. M. Redwing, Diffusion-controlled epitaxy of large area coalesced $WSe_2$ monolayers on sapphire. *Nano Lett.* **18**, 1049-1056 (2018).
45. P. Zheng, W. Wei, Z. Liang, B. Qin, J. Tian, J. Wang, R. Qiao, Y. Ren, J. Chen, C. Huang, X. Zhou, G. Zhang, Z. Tang, D. Yu, F. Ding, K. Liu, X. Xu , Universal epitaxy of non-centrosymmetric two-dimensional single-crystal metal dichalcogenides. *Nat Commun*. **14**, 592 (2023).
46. J. VandeVondele, M. Krack, F. Mohamed, M. Parrinello, T. Chassaing, J. Hutter, Quickstep: fast and accurate density functional calculations using a mixed Gaussian and plane waves approach. *Comput. Phys. Commun.* **167**, 103-128 (2005).
47. S. Goedecker, M. Teter, J. Hutter, Separable dual-space Gaussian pseudopotentials. *Phys. Rev. B* **54**, 1703-1710 (1996).

48. V. Vande, J. Hutter, Gaussian basis sets for accurate calculations on molecular systems in gas and condensed phases. *J. Chem. Phys.* **127**, 114105-114109 (2007).
49. J. Perdew, K. Burke, M. Ernzerhof, Generalized gradient approximation made simple. *Phys. Rev. Lett.* **77**, 3865-3868 (1996).
50. S. Grimme, J. Antony, S. Ehrlich, H. Krieg, A consistent and accurate ab initio parametrization of density functional dispersion correction (DFT-D) for the 94 elements H-Pu. *J. Chem. Phys.* **132**, 154104 (2010).
51. R. I. Eglitis, D. Vanderbilt, First-principles calculations of atomic and electronic structure of $SrTiO_3$ (001) and (011) surfaces. *Phys. Rev. B* **77**, 195408 (2008).
52. T. Lu, F. Chen, Multiwfn: A Multifunctional Wavefunction Analyzer, *J. Comput. Chem.* **33**, 580-592 (2012).
53. T. Lu, A comprehensive electron wavefunction analysis toolbox for chemists, Multiwfn, *J. Chem. Phys.* **161**, 082503 (2024).
54. J. Neugebauer, M. Scheffler, Adsorbate-substrate and adsorbate-adsorbate interactions of Na and K adlayers on Al(111). *Phys. Rev. B*, **46**, 16067 (1992).